\documentstyle[12pt,epsf]{article}

\renewcommand{\bar}[1]{\overline{#1}}
\newcommand{\hsim}[1]{\ \ \mathrel{\rlap{\lower-4pt\hbox{$\sim$}}
                    \hskip-3pt\hbox{$#1$}}\,}
\newcommand{\rtop} {\buildrel \Rightarrow \over {_{x\rightarrow 1}}}
\newcommand{\eqx} {\buildrel = \over {_{x \to 1}}}
\newcommand{\lsim} {\buildrel < \over {_\sim}}
\newcommand{\gsim} {\buildrel > \over {_\sim}}
\newcommand{\longvec}[1]{\overrightarrow{\!\!#1}}

\newcommand{\ket}[1]{\,\left|\,{#1}\right\rangle}
\newcommand{\M}{{\cal M}}

\begin{document}

\begin{flushright}
SLAC--PUB--8474\\
June 2000
\end{flushright}
\bigskip\bigskip

\thispagestyle{empty}

\vfill

\boldmath
{\centerline{\Large\bf Dynamical Higher-Twist and High x
Phenomena:}
\unboldmath
\centerline{\Large\bf A Window to Quark-Quark Correlations in QCD
    \footnote{\baselineskip=14pt
     Work supported by the Department of Energy, contract
     DE--AC03--76SF00515.}}
\vspace{22pt}
  \centerline{\bf Stanley J. Brodsky}
\vspace{8pt}
  \centerline{\it Stanford Linear Accelerator Center}
  \centerline{\it Stanford University, Stanford, California 94309}
  \centerline{e-mail: sjbth@slac.stanford.edu}
\vspace*{0.9cm}

\vfill
\begin{center}

Invited talk presented at the \\
``Workshop on Nucleon Structure in the High x-Bjorken Region
(HiX2000)''
\\ Temple University, Philadelphia, Pennsylvania \\
March 30--April 1, 2000
\end{center}
\vfill
\newpage

\begin{center}

\abstract{
Measurements of the power-law corrections to Bjorken
scaling and the behavior of structure
functions in the highly stressed $x_{bj} \to 1$ regime of
electroproduction can lead to new information on the
quark-quark correlations controlling the nucleon wavefunction
at far-off-shell kinematics.  Electroproduction on nuclei
at $ A > x_{bj} > 1$ are sensitive
to hidden-color components of the nuclear wavefunction.  A distinctive
dynamical higher-twist
${\cal O}(1/Q^2)$ correction, which is dynamically enhanced at high
$x_{bj}$,  can arise from the interference of amplitudes where
the lepton scatters from two different valence quarks of the target.
Measurements of the parity-violating left-right asymmetry $A_{LR}$ in
elastic and inelastic polarized electron scattering at large
$x_{bj}$ can confirm the structure of the quark-quark correlations
and other QCD physics at the amplitude level.}
\end{center}

\section{Introduction}

A fundamental question in QCD is the non-perturbative structure of hadrons
at the amplitude level---not just the single-particle flavor,
momentum, and helicity distributions of the quark constituents,  but also
the multi-quark, gluonic, and hidden-color correlations intrinsic to
hadronic and nuclear wavefunctions.  As I shall discuss here, detailed
measurements of the power-law corrections to Bjorken scaling and the
behavior of structure functions in the highly stressed $x_{bj} \to 1$
regime of electroproduction can lead to important new information on the
dynamical mechanisms and the underlying quark-quark correlations of the
target wavefunction.  In the case of light-nuclei, one can obtain
sensitivity to hidden-color components of the nuclear wavefunction from
measurements beyond the nucleon kinematic domain.  Measurements of the
parity-violating left-right asymmetry in the elastic and inelastic
scattering of polarized electrons can add important checks on the QCD
mechanisms underlying dynamical higher twist effects.

The $n-$parton amplitudes which interpolate between a hadron $H$ and
its quark and gluon degrees of freedom in QCD are the light-cone Fock
wavefunctions
$\psi_{n/H}(x_i,{\vec k_{\perp i}},\lambda_i).$
Formally, the light-cone expansion is constructed by quantizing QCD at
fixed light-cone time
\cite{Dirac:1949cp} $\tau = t + z/c$ and forming the invariant light-cone
Hamiltonian: $ H^{QCD}_{LC} = P^+ P^- - {\vec P_\perp}^2$ where
$P^\pm = P^0 \pm P^z$ \cite{PinskyPauli}.  The operator
$P^- = i {d\over d\tau}$ generates light-cone time translations.
The
momentum
$P^+$ and
$\vec P_\perp$ operators are independent of
the interactions.
The eigen-spectrum of the $ H^{QCD}_{LC}$ yields the entire
mass spectrum of color-singlet hadron states in QCD, together with
their respective light-cone wavefunctions.
For example, the
proton state satisfies:
$ H^{QCD}_{LC} \ket{\psi_p} = M^2_p \ket{\psi_p}$.

The projection of
the proton's eigensolution $\ket{\psi_p}$ on the color-singlet
$B = 1$, $Q = 1$ eigenstates $\{\ket{n} \}$
of the free Hamiltonian $ H^{QCD}_{LC}(g = 0)$ gives the
light-cone Fock expansion:
$ \left\vert \psi_p(P^+, {\vec P_\perp} )\right> = \sum_n \psi_n(x_i,
{\vec k_{\perp i}},
\lambda_i) \left\vert n;
x_i P^+, x_i {\vec P_\perp} + {\vec k_{\perp i}}, \lambda_i\right >$.
The light-cone momentum fractions of the constituents,
$x_i = k^+_i/P^+$ with $\sum^n_{i=1} x_i = 1,$ and the transverse
momenta ${\vec k_{\perp i}}$ with
$\sum^n_{i=1} {\vec k_{\perp i}} = {\vec 0_\perp}$ appear as
the momentum
coordinates of the light-cone Fock wavefunctions.  The actual physical
transverse momenta are
${\vec p_{\perp i}} = x_i {\vec P_\perp} + {\vec k_{\perp i}}.$ The
$\lambda_i$ label the light-cone spin $S^z$ projections of the quarks and
gluons along the $z$ direction.  The physical gluon
polarization vectors
$\epsilon^\mu(k,\ \lambda = \pm 1)$ are specified in light-cone
gauge by the conditions $k \cdot \epsilon = 0,\ \eta \cdot \epsilon =
\epsilon^+ = 0.$ The relative orbital and spin projections in
each Fock state sum to the $J_z$ of the hadron \cite{Brodsky:2000ii}. The
light-cone Hamiltonian
formalism thus provides a relativistic description of hadrons as
 many-particle systems of fluctuating parton number.

The LC wavefunctions $\psi_{n/H}(x_i,\vec
k_{\perp i},\lambda_i)$ are universal, process independent, and thus
control all hadronic reactions.  In the case of deep inelastic scattering,
one needs to evaluate the imaginary part of the virtual Compton
amplitude ${\cal M}[\gamma^*(q) p \to \gamma^* (q) p].$ The simplest
frame choice for electroproduction is
$q^+ = 0, q_\perp^2 = Q^2= - q^2,  q^- = {2 q\cdot p / P^+},
p^+ = P^+, p_\perp = 0_\perp, p^- = {M_p^2/ P^+}.  $ At leading twist,
soft final-state interactions are power-law suppressed in light-cone
gauge, so the calculation of the virtual Compton amplitude reduces to the
evaluation of matrix elements of the products of free quark currents of
the free quarks.  The absorptive amplitude imposes conservation of
light-cone energy:
$p^- +  q^- = \sum^n_i k^-_i$ for the $n-$particle Fock state.  In the
impulse approximation, where only one quark $q$ recoils against the
scattered lepton, this condition becomes
$$M_p^2 + 2 q\cdot p = {({\vec k}_{\perp q}+ {\vec q}_\perp)^2 + m_q^2
\over x_q} +
\sum_{i \ne q} {k_{\perp i}^2 + m_i^2 \over x_i}$$
If we neglect the transverse momenta $k^2_\perp$ relative to $Q^2$ in the
Bjorken limit
$Q^2
\to
\infty,$
$x_{bj} = {Q^2/ 2 q \cdot p}$ fixed,  we obtain the condition $x_q =
x_{bj}$; {\it i.e.}, the light-cone fraction $x_q= k^+/p^+$
of the struck quark is
kinematically fixed to be equal to the Bjorken ratio.
Contributions from high
$k^2_\perp = {\cal O}(Q^2)$ which originate from the perturbative QCD
radiative corrections to the struck quark line lead to the DGLAP
evolution equations.

Thus given the light-cone wavefunctions, one can compute \cite{BL80}
all of the leading twist helicity and
transversity distributions measured in polarized deep inelastic
lepton scattering \cite{Jaffe:1989up}.  For example,
the polarized quark distributions at resolution $\Lambda$ correspond to
\begin{eqnarray}
q_{\lambda_q/\Lambda_p}(x, \Lambda)
&=& \sum_{n,q_a}
\int\prod^n_{j=1} dx_j d^2 k_{\perp j}\sum_{\lambda_i}
\vert \psi^{(\Lambda)}_{n/H}(x_i,\vec k_{\perp i},\lambda_i)\vert^2
\\
&& \times \delta\left(1- \sum^n_i x_i\right) \delta^{(2)}
\left(\sum^n_i \vec k_{\perp i}\right)
\delta(x - x_q) \delta_{\lambda_a, \lambda_q}
\Theta(\Lambda^2 - {\cal M}^2_n)\ , \nonumber
\end{eqnarray}
where the sum is over all quarks $q_a$ which match the quantum
numbers, light-cone momentum fraction $x,$ and helicity of the struck
quark.  Similarly, the transversity distributions and
off-diagonal helicity convolutions are defined as a density matrix of the
light-cone wavefunctions.  This defines the LC
factorization scheme\cite {BL80} where the
invariant mass squared ${\cal M}^2_n = \sum_{i = 1}^n {(k_{\perp i}^2 +
m_i^2 )/ x_i}$ of the $n$ partons of the light-cone wavefunctions are
limited to $ {\cal M}^2_n < \Lambda^2$

The
light-cone wavefunctions also specify the multi-quark and gluon
correlations of the hadron.  For example,  the distribution of spectator
particles in the final state which could be measured in the proton
fragmentation region in deep inelastic scattering at an electron-proton
collider are in principle encoded in the light-cone wavefunctions.

There are many sources of power-law corrections to the standard
leading twist formula for deep inelastic structure functions.
Higher-twist corrections arise from QCD radiative corrections
(renormalons), final-state interactions, finite target mass effects
\cite{Nachtmann:1973mr}, the constituent masses, and their transverse
momenta $k_\perp.$ A derivation of some of these corrections is given in
Ref. \cite {Brodsky:1979gy}.  Despite the many sources of power-law
corrections to the deep inelastic cross section, certain
types of dynamical contributions stand out at large $x_{bj}$ since they
arise from compact, highly-correlated fluctuations of the proton
wavefunction.  As I will discuss in Section 3, there are particularly
interesting dynamical ${\cal O}(1/Q^2)$ corrections which occur from the
{\it interference} of quark currents; {\it i.e.}, contributions which involve
leptons scattering from two different quarks.

\boldmath
\section{Structure Functions at High x}
\unboldmath

The impulse approximation for inelastic
lepton proton scattering is not valid for calculations of
structure functions at fixed $Q^2$ and large
$x\sim 1$.  For example, as $x\rightarrow 1$, the struck quark becomes
far-off shell and spacelike; its Feynman virtuality is
\begin{equation} k^2_F = x\left[ M^2_p -\frac{\M^2_s+k^2_\perp}{1-x} -
\frac{k^2_\perp}{x}\right] \Rightarrow - \infty \ .
\label{eq:ab}
\end{equation}
Here $\M^2_s$ is the invariant mass of the spectator
system after the struck quark is removed, and $\longvec k_\perp$ is the
transverse momentum of the struck quark.  In the language of light-cone
perturbation theory, the light-cone wavefunction is evaluated far from
its light-cone energy shell; in particular, the identification
$x= k^+/p^+$ will break down at $x
\to 1$ since the spectator light cone momentum fractions $x_i$ are
all forced to be small.  The spectator terms in the light-cone energy
conservation equation $p^- +  q^- = \sum^n_i k^-_i$ thus cannot be
ignored.

Thus the regime $x\rightarrow 1$ probes a highly stressed far
off-shell configuration of the proton wavefunction where the
struck quark has all of the proton's light-cone momentum and all the
spectator quarks and gluons are left with
negligible light-cone momentum fraction.  This regime clearly is
highly sensitive to the inter-particle correlations of the proton's
wavefunction; {\it i.e.} the detailed dynamics which allows all of the
proton's momentum to be transferred to just one quark.  In fact, in this
far-off-shell domain we can use PQCD to calculate the $x\rightarrow 1$
dependence of the structure functions \cite{BL80} by iterating the
equations of motion.  Only the lowest valence light-cone Fock state
contributes since there are the fewest number of spectators to stop.  To
leading order in
$\alpha_s(k^2_F)$, one can calculate the end-point dependence of
$F_2(x,Q^2)$ via two hard gluon exchanges between the three valence
quarks.  A typical perturbative QCD contribution is illustrated in Fig.
\ref{fig:8548A3}.  The
result is

\begin{figure}[htb]
\begin{center}
\leavevmode
{\epsfxsize=2.5in\epsfbox{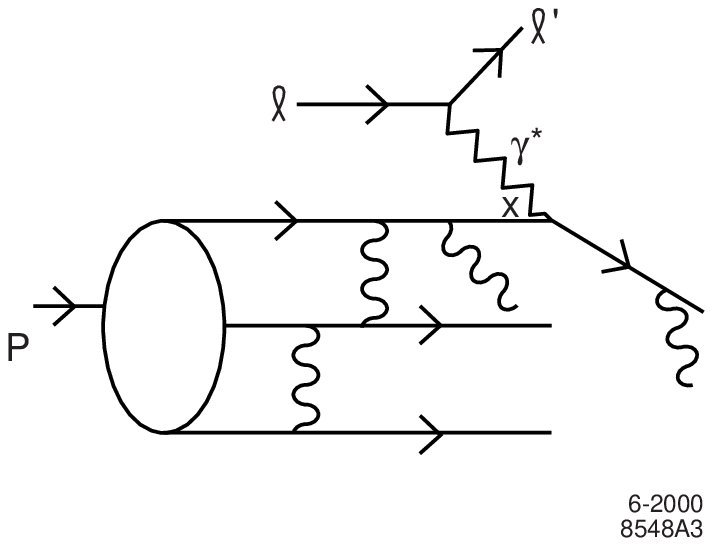}}
\parbox{5.2in}{\caption[*]{Perturbative QCD two-gluon-exchange
mechanism dominating
nucleon structure functions at $x \to 1$.  The dominant helicity of the
struck quark is parallel to that of the nucleon.  Gluon radiation from the
struck quark leads to DGLAP evolution if
$Q^2 > |k^2_f|$, the virtuality of the struck quark.  }}
\label{fig:8548A3}
\end{center}
\end{figure}
\begin{equation}
q_{\uparrow/\uparrow}(x,Q^2)     \hsim{_{x\rightarrow 1}} (1-x)^3\qquad
q_{\downarrow/\uparrow}(x,Q^2) \hsim{_{x\rightarrow 1}} (1-x)^5
\label{eq:ac}
\end{equation}
{\it i.e.}: it is much more probable that the leading quark has the same
helicity as that of
the proton:
\begin{equation}
\frac{u_{\downarrow/\uparrow}(x,Q^2)}{u_{\uparrow/\uparrow}(x,Q^2)}
\hsim{_{x\rightarrow 1}} (1-x)^2 \qquad
\frac{d_{\downarrow/\uparrow}(x,Q^2)}{d_{\uparrow/\uparrow}(x,Q^2)}
\hsim{_{x\rightarrow 1}} (1-x)^2 \ .
\label{eq:ad}
\end{equation}
If one assumes $SU(6)$-flavor symmetry,
then there are 5 times more $u\uparrow$ quarks than
$d\uparrow$ quarks in the proton:
\begin{equation}
\frac{u_\uparrow(x,Q^2)}{d_\uparrow(x,Q^2)} \rtop 5 .
\label{eq:ae}
\end{equation}
This also implies the famous Farrar-Jackson
prediction \cite{Farrar:1975yb}
\begin{equation}
\frac{F_{2n}(x,Q^2)}{F_{2p}(x,Q^2)} \rtop
\frac{5\cdot \left(\frac{1}{3}\right)^2 + \left(\frac{2}{3}\right)^2}
{5\cdot\left(\frac{2}{3}\right)^2 + \left(\frac{1}{3}\right)^2} =
\frac{3}{7} .
\label{eq:af}
\end{equation}
In the case of gluons, the leading PQCD prediction is
\begin{equation}
g_{\uparrow/\uparrow}(x,Q^2) \sim (1-x)^4 \qquad
g_{\downarrow/\uparrow}(x,Q^2) \sim (1-x)^6
\label{eq:ag}
\end{equation}
{\it i.e.}: the gluon polarization becomes strongly aligned with that of
the proton when the gluon takes all of the proton's light-cone momentum.
One also expects dominance of the helicity-aligned strange,
$\bar u$, and $\bar d$ distributions at $x \to 1$.

Useful phenomenological models of the input spin-dependent structure
functions $q_{\lambda/\lambda_p}(x,Q^2_0)$ can be designed which
incorporate the PQCD-predicted power laws at
$x\rightarrow 1$ and isospin-singlet $1/x$
Pomeron and isospin-nonsinglet
$1/\sqrt x$ Reggeon behavior at small $x$ \cite{Brodsky:1995kg}.
Such forms match well with the MRS parameterizations
of the data \cite{Martin:1993zi}.  There are a wide range of QCD flavor and
helicity tests of these predictions which could be carried out at a 12
GeV facility.  For example, a simple model for the polarized gluon
distribution in the proton is \cite{Brodsky:1990db,Brodsky:1995kg}
\begin{eqnarray}
g_{\uparrow/\uparrow}(x,Q^2) &=& A \frac{(1-x)^4}{x} \nonumber\\[1ex]
g_{\downarrow/\uparrow}(x,Q^2) &=& A \frac{(1-x)^6}{x} \\[1ex]
\Delta g(x,Q^2) &=& A(1-x)^4 (2-x) \nonumber \ .
\label{eq:ah}
\end{eqnarray}
If the momentum carried by gluons is
\begin{eqnarray}
\int^1_0 dx \, x(g_{\uparrow/\uparrow}(x) + g_{\downarrow/\uparrow(x)}) =
{1\over 2},
\label{eq:b1}
\end{eqnarray}
then
$A = 35/24$, and $\Delta g = 77/144 \cong 0.54$.
These predictions are expected to be applicable at the starting scale for
PQCD evolution; {\it i.e.} $Q^2 \lsim$ 2 GeV$^2$.

It is also interesting to measure inelastic lepton-nucleus scattering
at $1 < x_{bj} < A$, beyond the kinematic domain accessible on a single
nucleon target.  The nuclear light-cone momentum must be transferred to a
single quark, requiring quark-quark correlations between
quarks of different nucleons in a compact, far-off-shell regime.  The
nuclear wavefunction contains hidden-color components distinct from a
convolution of separate color-singlet nucleon wavefunctions.  In fact, at
very short distances, the light-cone distribution amplitude of a deuteron
must involve asymptotically into a state which
is $80$\% hidden color \cite{Brodsky:1983vf}.

How does DGLAP evolution affect the $x \rightarrow 1$ dependence?
Usually one expects that structure functions are strongly suppressed at
large
$x$ because of the momentum lost by gluon radiation:  the predicted
change of the power law behavior at large $x$ is \cite{Gribov:1972ri}
\begin{equation}
\frac{F_2(x,Q^2)}{F_2(x,Q^2_0)} \eqx (1-x)^{\zeta(Q^2,Q^2_0)}
\label{eq:ai}
\end{equation}
where
\begin{equation}
\zeta(q^2,Q^2_0) = \frac{1}{4\pi} \int^{Q^2}_{Q^2_0}
\frac{d\ell^2}{\ell^2}\, \alpha_s(\ell^2) \ .
\label{eq:aj}
\end{equation}
Because of asymptotic freedom, this implies a $\log \log Q^2$ increase in
the effective power $\zeta(Q^2,Q^2_0)$.  However, this derivation assumes
that the struck quark is on its mass shell.  The off-shell effect is
profound, greatly reducing the PQCD
radiation \cite{Brodsky:1979gy, Lepage:1982gd}.
We can take into account the main effect of the struck quark virtuality by
modifying the propagator in Eq. (\ref{eq:aj}):
\begin{equation}
\zeta(Q^2,Q^2_0) = \frac{1}{4\pi} \int^{Q^2}_{Q^2_0}
\frac{d\ell^2}{\ell^2+|k^2_f|}\,
\alpha_s(\ell^2) .
\label{eq:ak}
\end{equation}
Thus at large $x$, there is effectively no DGLAP evolution until
$Q^2\gsim |k^2_f|$!  One can also see that DGLAP
evolution at large
$x$ at fixed $Q^2$ must be suppressed in order to have
duality at fixed
$W^2 = Q^2(1-x_{bj})/x_{bj}$ between the inclusive electroproduction and
exclusive resonance contributions \cite{BL80}.

\section{Higher-Twist Signals in Electroproduction}

It is an empirical fact that conventional leading twist contributions
cannot account for the measured $e p
\rightarrow e X$ and $e d \rightarrow e X$ structure functions at $x
\gsim$ 0.4 and $Q^2 \lsim$ 5 GeV$^2$.  Fits to the
data \cite{Virchaux:1992jc,Amaudruz:1992nw} require an additional
component which scales as
$1/Q^2$ relative to the leading twist contributions and rises
rapidly with
$x$.  The excess contribution can be parameterized in the form
\begin{figure}[htb]
\begin{center}
\leavevmode
{\epsfxsize=3in\epsfbox{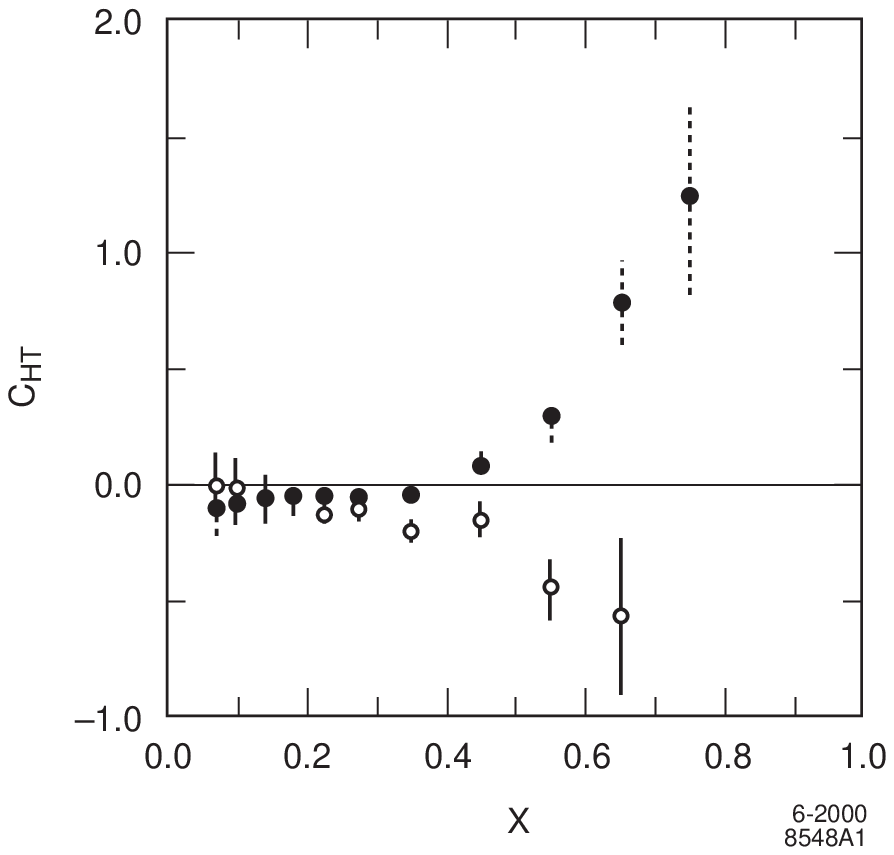}}
\parbox{5.2in}{\caption[*]{Higher-twist coefficients $C_{HT}(x)$ [in
GeV$^2$ units] for
inelastic lepton scattering on proton target (solid points) and the
difference $C^p_{HT}(x)-C^n_{HT}(x)$ for proton minus neutron targets
(open circles),  from
Refs. \cite{Virchaux:1992jc, Amaudruz:1992nw}.  The data compilation
is taken from Ref. \cite {Souder:2000}.}}
\label{fig:8548A1}
\end{center}
\end{figure}

\begin{equation}
F_{2p,n}(x,Q^2) = F^0_{2p,n}(x,Q^2)
\left[1+ \frac{c_{HT}^{p,n}(x)}{Q^2}\right]
\label{eq:al}
\end{equation}
where $F^0_{2p,n}$ is the leading twist contribution.  The functional
dependence of the higher-twist term $C_{HT}^{p,n}(x)$ for proton and
proton-neutron targets is shown in Fig. \ref{fig:8548A1}.  A rough fit is
\begin{equation}
c_{HT}^p(x) \cong \left[\frac{0.3\ GeV}{1-x}\right]^2
\qquad
c_{HT}^n(x) \cong 2 c_{HT}^p(x) \ ;
\label{eq:am}
\end{equation}
{\it i.e.}:
the higher-twist effect relative to the leading twist contribution for
the neutron is stronger than that of the proton.

A possible source of higher-twist effects in PQCD
is ``renormalons'' \cite{Beneke:1999ui,Maul:1997rz}.  This
contribution to the deep inelastic lepton-hadron cross section reflects
a divergent $\beta^n_0\, n$! growth of the PQCD series for hard
radiative corrections to deep inelastic scattering evolution at
high orders in $\alpha^n_s(Q^2)$.  The factorial growth arises from the
integration over the QCD running coupling; {\it i.e.}, the summation
of the reducible multi-bubble loop-diagrams in the gluon propagator.
The net
effect is to correct the leading twist predictions by a power-law
suppressed $1/Q^2(1-x)$ contribution.
Alternatively, one can proceed using the BLM method \cite{Brodsky:1983gc}:
one first
identifies the conformal coefficients \cite{Brodsky:2000cr} of the
PQCD series; by definition these are independent of the
$\beta-$function and are hence devoid of the
$\beta^n_0\, n!$ growth.  The scale of the running coupling is
set by requiring that all of the
$\beta$-dependence resides in $\alpha_s(Q^{*2}).$
The resulting scale
$(Q^{*2}) \propto (1-x)Q^2$ can also be understood as the mean value
of the argument of the running coupling $\alpha_s(k^2)$ in the Feynman
loop integration.

However, the renormalon contribution cannot account
for the observed higher-twist contribution shown in Fig. \ref{fig:8548A1}
since it is proportional to the leading-twist prediction, {\it i.e.}:
$c_{HT ren}^p(x) = c_{HT ren }^n(x) .$
Thus it is apparent from the data that there must be a dynamical origin
for the observed $C_{HT}(x)/Q^2$ contribution.  In fact, dynamical
higher-twist terms naturally arise from multi-parton correlations.  For
example, if the electron recoils against 1, 2, or 3 quarks, one obtains
a series of higher-twist contributions of ascending order in $1/Q^2$.
\begin{eqnarray}
\sigma_T &\sim& \frac{(1-x)^3}{Q^2)} \qquad e q \rightarrow eq \nonumber
\\[1ex]
\sigma_L &\sim& \frac{(1-x)}{(Q^2)^3} \qquad e qq \rightarrow e qq \\[1ex]
\sigma_T &\sim& \frac{1}{(1-x)} \left(\frac{1}{Q^2}\right)^3 \quad e qqq
\rightarrow
e qqq \nonumber
\label{eq:ao}
\end{eqnarray}
where the extra $1/Q^2$ fall-off reflects the form factor squared of the
$(qq)$ or
$(qqq)$ systems, and the enhancement at $x \rightarrow 1$ reflects the
fact that the $(qq)$ and $(qqq)$ composites carry increasing fractions of
the proton light-cone momentum.  The dominance of $\sigma_L$ for $eqq
\rightarrow eqq$ reflects the bosonic coupling of the composite di-quark.
Each of the contributions satisfy Bloom-Gilman
duality \cite{Bloom:1970xb} at fixed $W^2$. The multi-parton subprocesses
are suppressed by powers of
$1/Q^2$ but enhanced at large $x$ since more of the momentum of the
target proton is fed into the hard subprocess; {\it i.e.}, there are fewer
spectators to stop.  The general rule is
$$F_2(x,Q^2)\propto {(1-x)^{2 n_{spect} - 1 + 2 \Delta h}\over
Q^{n_{active}-4}}$$ where $n$ is the number of partons or other quanta
participating in the hard subprocess and $\Delta h$ is the difference in
helicity between the active partons and the
target \cite{Blankenbecler:1978vk}.

\vspace{.5cm}
\begin{figure}[htb]
\begin{center}
\leavevmode
{\epsfbox{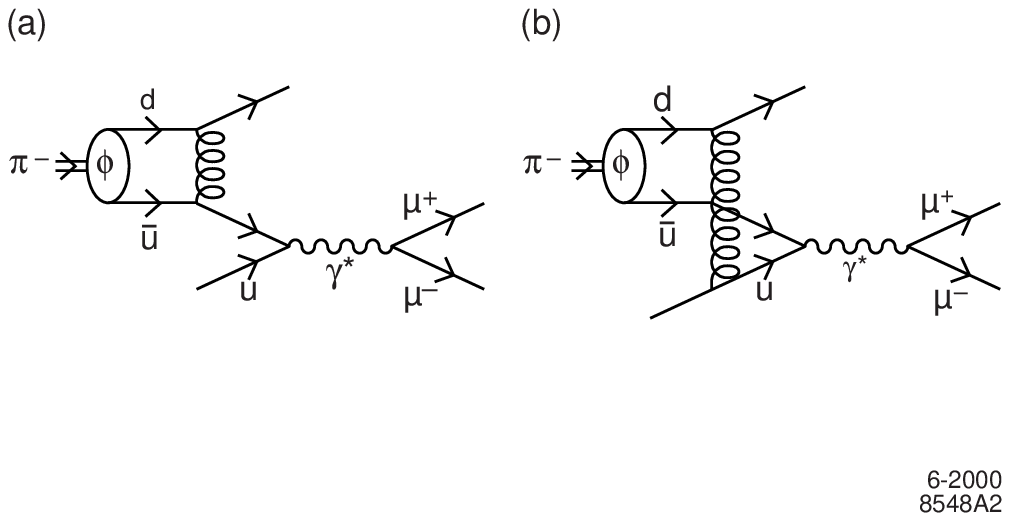}}
\parbox{5.2in}{\caption[*]{Higher-twist contribution to lepton pair
production in $\pi
N$ scattering.  The dynamics at large $x_F$ requires both constituents of
the projectile meson to be involved in the hard subprocess.  From Ref.
\cite{Brandenburg:1994wf}.}}
\label{fig:8548A2}
\end{center}
\end{figure}

It is well-known that higher-twist, power-law suppressed corrections to
hard inclusive cross sections can be a signature of correlation
effects involving two or more valence quarks of a hadron.  For example,
the lepton angular dependence of the leading-twist PQCD prediction for
Drell Yan lepton pair production
$d\sigma (\pi A \to \ell^+ \ell^- X)/d \Omega$ is $1 + \cos^2
\theta_{cm}$. The data of Ref. \cite{Guanziroli:1988rp,Conway:1989fs}
however shows the onset of $\sin^2 \theta_{cm}$ dependence at large
$x_F$.  This signals the presence of multiparton-induced subprocesses
such as
$(\bar q q) q \to \gamma^*(Q^2) q \to \ell^+ \ell^- q$ \cite{Berger:1979du}.
See Fig. \ref{fig:8548A2}.
Such reactions
produce longitudinally-polarized virtual photons with a $\sin^2
\theta_{cm}$ lepton pair angular dependence in contrast to the
transversally polarized Drell-Yan pairs produced from the
${\bar q } q \to \gamma^*(Q^2) \to \ell^+ \ell^-$ subprocess.
The penalty for utilizing the two correlated partons in the pion
wavefunction is an extra suppression factor $1/ R^2 Q^2 (1-x_F)^2$
where
$R$ is the characteristic interquark transverse separation between the
valence quarks in the incoming meson.  The origin of the $1/R^2 Q^2$
scaling is similar to that of the photon to meson transition form factor
in the exclusive reaction $\ell \gamma \to \ell (\bar q q) \to
\pi^0$ \cite{BL80}.  The scale $1/R$ can be related to the pion decay
constant $f_\pi$ which normalizes the pion distribution
amplitude \cite{BL80}.  At fixed
$Q^2$ the higher-twist process can actually dominate as $x_F \to 1$ since
all of the incoming momentum of the pion is transferred to the
subprocess.  The correlated subprocess
$(\bar q q) q
\to \gamma^*(Q^2) q \to \ell^+ \ell^- q$ also leads to the prediction of
$\sin^2\theta \cos 2 \phi$ and $\sin 2 \theta \cos \phi$
terms in the angular distribution \cite{Brandenburg:1994wf}, effects which
are clearly apparent in the data
\cite{Guanziroli:1988rp,Conway:1989fs}.

Another important example of dynamical higher-twist effects is the
reaction $\pi A
\to J/\psi X$ which is observed to produce longitudinally-polarized
$J/\psi's$ at large
$x_F$ \cite{Biino:1987qu}.  Again this effect can be attributed to highly
correlated multi-parton subprocesses such as $\bar q q g \to \bar c c
\bar q q$ where both valence quarks of the incident pion must be involved
in the hard subprocess in order to produce the charmed quark pair with
nearly all of the incident momentum of the incoming
meson \cite{Vanttinen:1995sd}.  Similarly, charm production at
threshold requires that all of the momentum of the target nucleon
be transferred to the charm quarks.  In the $\gamma p \rightarrow
c\overline{c}p$ reaction near threshold,  all the partons have to
transfer their energy to the charm quarks within their reaction time
$1/m_c$, and must be within this transverse distance from the
$c\overline{c}$ and from each other.
Hence only compact Fock states of the
target nucleon or nucleus with a radius equal to the Compton wavelength of
the heavy quark, can contribute to charm production at threshold.
Equivalently we can interpret the multi-connected charm quarks as
intrinsic charm Fock states which are kinematically favored to have
large momentum fractions  \cite{Brodsky:1980pb}.
The experimental evidence for intrinsic charm is discussed
in Ref. \cite{Harris:1996jx}

Near-threshold charm production also probes the $x\simeq 1$ configurations
in the target wavefunction; the spectator partons carry a vanishing
fraction
$x\simeq 0$ of the target momentum.  This implies that the production
rate behaves near $x\rightarrow 1$ approximately as $(1-x)^{2n_s-1}$
where
$n_s$ is the number of spectators required to stop.  Including spin
factors, we can identify three different gluonic components of the
photoproduction cross-section:
\begin{itemize}
\item The usual one-gluon $(1-x)^4$ distribution for leading twist
photon-gluon fusion $\gamma g\rightarrow c \overline{c}$, which leaves
two quarks spectators;
\item Two correlated gluons emitted from the proton with a net
distribution \hfill\break ${(1-x)^2}/{R^2{\cal{M}}^2}$ for $\gamma gg
\rightarrow
c\overline{c}$, leaving one quark spectator;
\item Three correlated gluons emitted from the proton with a net
distribution \hfill\break $
{(1-x)^0}/{R^4{\cal{M}}^4}$ for $\gamma ggg \rightarrow
c\overline{c}$, leaving no quark spectators.
\end{itemize} Here $x \approx {\cal{M}}^2/(s-m^2)$ and $\cal{M}$ is the
mass of the $c\overline{c}$ pair.  The relative weight of the
multiply-connected terms is controlled by the inter-quark separation
$R\simeq 1/m_c$.  The extra powers of $1/\cal{M}$ arise from the
power-law fall-off of the higher-twist hard subprocesses \cite{BHL}.

The correlations between valence quarks can also have an important effect
in deep inelastic scattering, particularly at large $x_{bj}= {Q^2/ 2 p
\cdot q}$.   As noted above,
one expects a sum of contributions to the
deep inelastic cross section scaling nominally as
$$F_2(x,Q^2) = A(1-x)^3 + B{(1-x)^2\over Q^4} + C{(1-x)^{-1}\over Q^8}$$
corresponding to the subprocesses $\ell q \to \ell q$, $\ell ( q q) \to
\ell (q q)$, and $\ell ( q qq ) \to \ell (q q q)$.
However, the above classification of terms in $F_2(x,Q^2)$ neglects what
may be the most significant and interesting higher-twist contribution to
deep inelastic scattering: the interference contributions.  Let us
consider the contribution to DIS due to the interference of the amplitude
where the lepton scatters on one quark with the amplitude where the lepton
scatters on another quark.  See Fig. \ref{fig:8548A4}.  One might think
such contributions are assumed to be negligible since the hard
subprocesses seem to lead to different non-interfering final states.
Actually these contributions can interfere if the struck quarks have high
internal momentum in the initial state or if they exchange large momenta
in the final state.  In either case, the apparently different final states
can overlap.  An insightful nuclear physics analog has been discussed by
Drell \cite{Drell:1992rx}.

\vspace{.5cm}
\begin{figure}[htb]
\begin{center}
\leavevmode
{\epsfbox{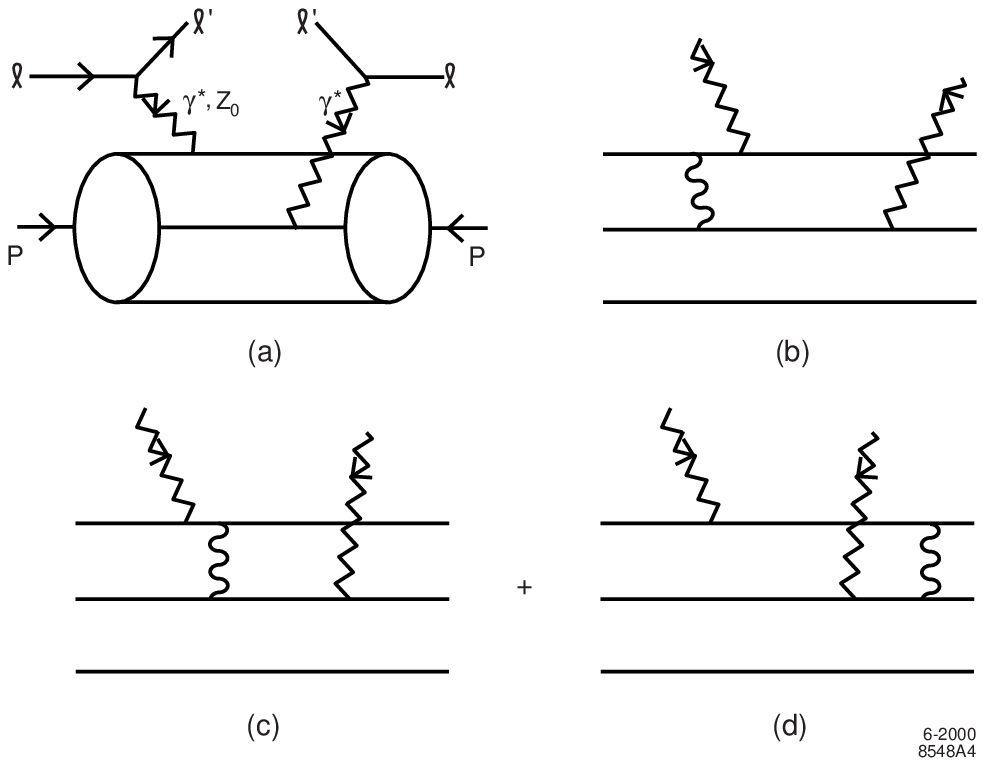}}
\parbox{5.2in}{\caption[*]{(a) Twist-four contribution to inelastic lepton
scattering where the lepton scatters on different quarks.  The
interference of
$\gamma^*$ and
$Z^0$ exchange contributions leads to parity and charge-conjugation
violation of the higher-twist contribution.  (b-d) The leading-order
${\cal O} (\alpha_s / Q^2 R^2)$ perturbative QCD gluon-exchange
contributions.  The higher-twist contribution to the structure function is
obtained by a convolution of the nucleon light-cone wavefunctions with the
$\gamma^* (q q) \to \gamma^* (q q)$ multi-quark amplitude.}}
\label{fig:8548A4}
\end{center}
\end{figure}

Let us consider the electroproduction subprocess
$\ell (q q) \to
\ell q q$ where the initial
$(q q)$ are collinear and have small invariant mass in the initial state
and the
$q q$ pair in the final state can have large invariant mass.  The lepton
can effectively scatter on either quark.  The nominal scaling of such
twist-four contributions is
$$F^{\rm interference}_2(x,Q^2) = \sum_{a \ne b}
e_a e_b {(1-x)^2\over R_{a b}^2 Q^2 }$$
where the factor of $1/R^2_{a b}$ reflects the inter-parton distance.
The interference terms are distinctive since, unlike renormalon
contributions, they do not track with the leading twist contributions.
The growth at high $x$ of the twist-four process reflects the fact that
the
$\ell (q q) \to
\ell q q$ subprocess incorporates the momentum of both quarks.  This
contribution must also play an important role in the physics of
Bloom-Gilman duality
\cite{Bloom:1970xb} since the interference contributions also appear in
square of the transition form factors.  The interference terms can
contribute to both
$F_L$ and $F_T$.  There is an extensive literature on higher-twist
contributions to the structure functions coming from such four-fermion
operators \cite{Capitani:1999ai,Edelmann:2000yp}.  They are also referred
to as ``cat ear" diagrams from their appearance in the virtual Compton
amplitude.

Let us suppose that the proton wavefunction is symmetric in the
coordinates of the three valence quarks.  If we sum over the
pairs of valence quarks, we obtain a vanishing contribution on a proton
target
$$\sum_{a \ne b} e_a e_b = (\sum_a e_a)^2 - \sum_a e_a^2 = 1 -
(4/9+4/9+1/9) = 0. $$
However, for the neutron
$$\sum_{a \ne b} e_a e_b = (\sum_a e_a)^2 - \sum_a e_a^2 = 0 -
(4/9+1/9+1/9) = 2/3.$$
Thus for symmetric nucleon wavefunctions the dynamical higher-twist
cross terms appear to be are zero in the proton and significant for the
neutron, deuteron, and nuclei!  This is a very distinctive effect; it
particularly motivates the empirical study of higher-twist effects using
the deuteron and nuclear targets.

In a more realistic treatment, one needs to take into account
correlation substructure.  For example, suppose that we
can approximate the nucleon wavefunctions as quark di-quark composites,
where the di-quark has $I=0$ and $J=0.$ Let us also suppose that
the inter-quark separation
$R_{a b}$ is smallest for the two quarks of the diquark composite.  In
this case we can approximate the full sum as a sum over the quark charges
of the $I= 0$ $u d$ diquark.  Then
$\sum_{a \ne b} e_a e_b = e_u e_d = -2/9$
for both the proton and neutron targets.  However, since it is
conventional to parameterize the higher-twist contribution as
a correction to the leading twist term.  Thus $C_{p,n}(x)$
is predicted to rise strongly at large $x$ and
$C_n(x) $ will be larger than $C_p(x)$ since the leading-twist
contribution to the neutron structure function
$F^{n}_2(x,Q^2)$ is significantly smaller than $F^{p}_2(x,Q^2)$.  These
predictions seem consistent with the empirical higher-twist contributions
to electroproduction extracted in Refs. \cite{Virchaux:1992jc} and
\cite{Amaudruz:1992nw}.  A simple test of the $I = 0$ diquark
higher-twist model  is the absence of twist-four contributions to the
combination of structure functions
$F^{d}_2(x,Q^2) - 2 F^{p}_2(x,Q^2)$

It is also
interesting to note that one can have interference between the
amplitude for lepton-quark scattering via photon exchange on one quark
with the amplitude for $Z^0$ exchange on another quark.  This implies a
distinctive parity-violating higher-twist contribution
$C^{PV}_{HT}(x)$ proportional to the product of electromagnetic and weak
quark charges
$\sum_{a \ne b} e^\gamma_a e^{Z^0}_b$.  Twist-four contributions of this
type have been in fact been modeled in Ref. \cite{Castorina:1985uw} for
structure function moments. However there is also the possibility of
high-$x$ enhancement.  In fact, the
$x$-dependence of $C^{PV}_{HT}(x)$ should be similar to the
parity-conserving contribution.

We can also use Bloom-Gilman duality to predict that the parity-violating
structure functions at large $x$ should average to the contributions of
the elastic and inelastic electroproduction channels when integrated over
similar ranges in $W^2$.  In fact, the parity-violating elastic form
factors can be predicted at large momentum transfer in
perturbative QCD  \cite{Brodsky:1981sx}.  Such measurements will provide
very interesting tests of the applicability of PQCD to
exclusive processes.
Thus as emphasized by
Souder \cite{Souder:2000}, the detailed measurement of the left-right
asymmetry $A_{LR}$ in polarized elastic and inelastic electron-proton and
polarized electron nucleus scattering at large
$x_{bj}$ can be a powerful illuminator of quark-quark correlations and
fundamental QCD physics at the amplitude level.

\section*{Acknowledgment}
I thank Paul Hoyer, Jean-Marc Laget, and Paul Souder
for helpful conversations.

\end{document}